\documentclass[epjST]{svjour}
\usepackage{graphics}
\usepackage{epstopdf}
\usepackage{subfigure}
\usepackage{amsmath}
\usepackage{amssymb}
\begin{document}
\title{Two-dimensional short-range interacting attractive and repulsive Fermi gases at zero temperature.}
\author{Gianluca Bertaina\thanks{\email{gianluca.bertaina@zoho.com}} }
\institute{Institute of Theoretical Physics, Ecole Polytechnique F\'{e}d\'{e}rale de Lausanne EPFL, CH-1015 Lausanne, Switzerland}
\abstract{
We study a two-dimensional two-component Fermi gas with attractive or repulsive short-range interactions at zero temperature.  We use Diffusion Monte Carlo with Fixed Node approximation in order to calculate the energy per particle and the opposite-spin pair distribution functions. We show the relevance of beyond mean field effects and verify the consistency of our approach by using Tan's Contact relations. 
} %end of abstract
\maketitle
\section{Introduction}
\label{intro}
Low dimensional configurations of degenerate Fermi and Bose gases are being object of many experimental and theoretical studies~\cite{RMP} as a means of simulating strongly correlated systems and as interesting systems per se, from a fundamental point of view. 

Many important areas of investigation on Fermi gases are being extended to two-dimensional (2D) configurations, which have already been pursued in the three-dimensional (3D) case, such as the BCS-BEC crossover in a superfluid gas with resonantly enhanced interactions and the possible onset of itinerant ferromagnetism in a gas with repulsive interactions. Other interesting phenomena that should be prominent in the 2D case are the Fulde-Ferrell-Larkin-Ovchinnikov superfluid state, the ultracold gases analogue of the quantum Hall effect in presence of non-abelian gauge fields and the long range correlations induced by dipolar interactions.

In particular, 2D ultracold Fermi gases have received a lot of attention in recent years, having been realized using highly anisotropic pancake-shaped potentials. Density profiles of the clouds have been measured using {\it in situ} imaging~\cite{Turlapov,Orel}; the single particle spectral function has been measured by means of rf spectroscopy \cite{Feld}, the presence of many-body polaronic states and their transition to molecular states have been characterized \cite{koschorreck}. 

On the theoretical side, the solution of the BCS equations for the 2D attractive Fermi gas has been first investigated in \cite{Miyake83} and later in \cite{Randeria}. The perturbative analysis of the 2D repulsive Fermi gas has been performed in \cite{Bloom75}. Recent experiments are in a regime where beyond mean-field contributions become relevant. In such a respect, many results are now available especially for the highly imbalanced case \cite{polaron}, both for the ground-state and for the so-called Upper Branch (UB). The 2D case is particularly challenging from the point of view of theory, being a marginal case in field theories, when the leading dependence of the observables on the coupling is non algebraic. Recently we have obtained the first determination using Quantum Monte Carlo methods of the equation of state at $T=0$ of a balanced homogeneous 2D Fermi gas in the BCS-BEC crossover \cite{bertaina}. Such an equation of state has been positively compared to experiments using the local density approximation \cite{Orel}, it has been used to discuss recent results on collective modes in a 2D Fermi gas \cite{collective} and to judge the temperature dependence of the Contact in 2D trapped gases \cite{Frohlich2012}.

In this article we report new results concerning the equation of state of the repulsive gas and its density-density correlation function. Moreover we emphasize the need of a consistent choice for the coupling constants and the coefficients of the beyond mean-field contributions to energy, both in the strongly interacting molecular regime and in the weakly interacting Fermi liquid regime. In Section \ref{sec:method} we introduce the model potentials and we discuss the trial nodal surfaces used in the Quantum Monte Carlo method; in Section \ref{sec:energy} we present the equation of state of the weakly attractive or repulsive gases, we show results for the Upper Branch of the attractive gas and we discuss the equation of state of the composite bosons in the molecular regime; finally in Section \ref{sec:tan} we discuss the extraction of the Contact from the density-density correlation function and we show results for the Contact of the repulsive gas.

\section{Method}
\label{sec:method}
We consider a homogeneous two-component Fermi gas in 2D described by the Hamiltonian
\begin{equation}
H=-\frac{\hbar^2}{2m}\left( \sum_{i=1}^{N_\uparrow}\nabla^2_i + \sum_{i^\prime=1}^{N_\downarrow}\nabla^2_{i^\prime}\right)
+\sum_{i,i^\prime}V(r_{ii^\prime}) \;,
\label{hamiltonian}
\end{equation}   
where $m$ denotes the mass of the particles, $i,j,...$ and $i^\prime,j^\prime,...$ label, respectively, spin-up and spin-down
particles and $N_\uparrow=N_\downarrow=N/2$, $N$ being the total number of atoms. We model the interspecies interatomic 
interactions using three different types of model potentials: an attractive square-well (SW) potential $V(r)=-V_0$ for $r<R_0$ ($V_0>0$), and $V(r)=0$ otherwise; a repulsive soft-disk (SD) potential $V(r)=V_1$ for $r<R_1$ ($V_1>0$), and $V(r)=0$ otherwise; and a hard-disk (HD) potential  $V(r)=\infty$ for $r<R_2$ and $V(r)=0$ otherwise. Due to the logarithmic dependence on energy of the phase shifts in 2D, different definitions of the scattering length have been used~\cite{Petrov03}; we define the scattering length $a_{2D}=R_2$ in the case of the HD potential, so that for the SD potential one gets $a_{2D}=R_1\;e^{-I_0(\kappa_1)/\kappa_1 I_1(\kappa)}$ and for the SW potential $a_{2D}=R_0\;e^{J_0(\kappa_0)/\kappa_0 J_1(\kappa_0)}$, where $J_{0(1)}$ and $I_{0(1)}$ are the Bessel and modified Bessel functions of first kind and $\kappa_0=\sqrt{V_0mR_0^2/\hbar^2}$, $\kappa_1=\sqrt{V_1mR_1^2/\hbar^2}$. In order to ensure the diluteness of the attractive gas we use $nR_0^2=10^{-6}$, where $n$ is the gas number density. The Fermi wave vector is defined as $k_F=\sqrt{2\pi n}$, and provides the energy scale $\varepsilon_F=\hbar^2 k_F^2/2m$. 
In order to assure universality in the repulsive case we check that the results for the HD potential and for the SD potential with different values of $R_1$ are compatible.
\begin{figure}
\centering
\subfigure[]{\resizebox{0.425\columnwidth}{!}{%
  \includegraphics{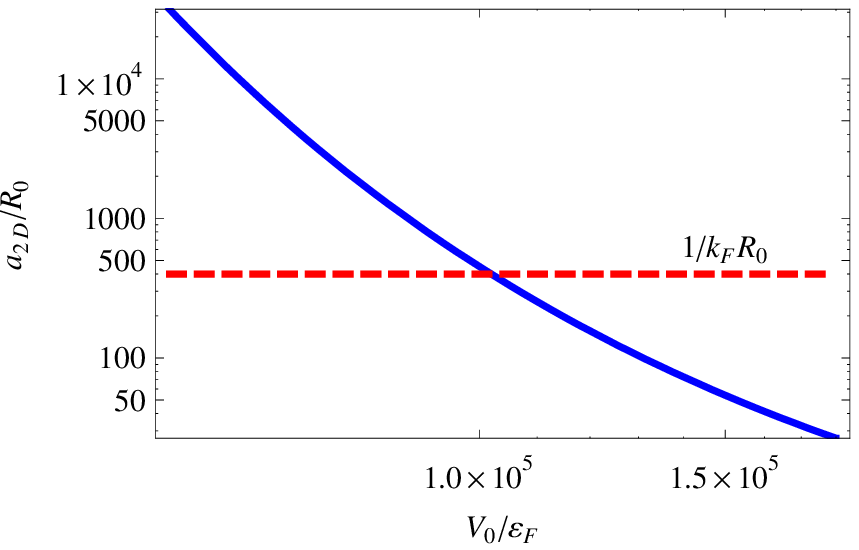} }\label{fig:a2dSW}}
\subfigure[]{\resizebox{0.40\columnwidth}{!}{%
  \includegraphics{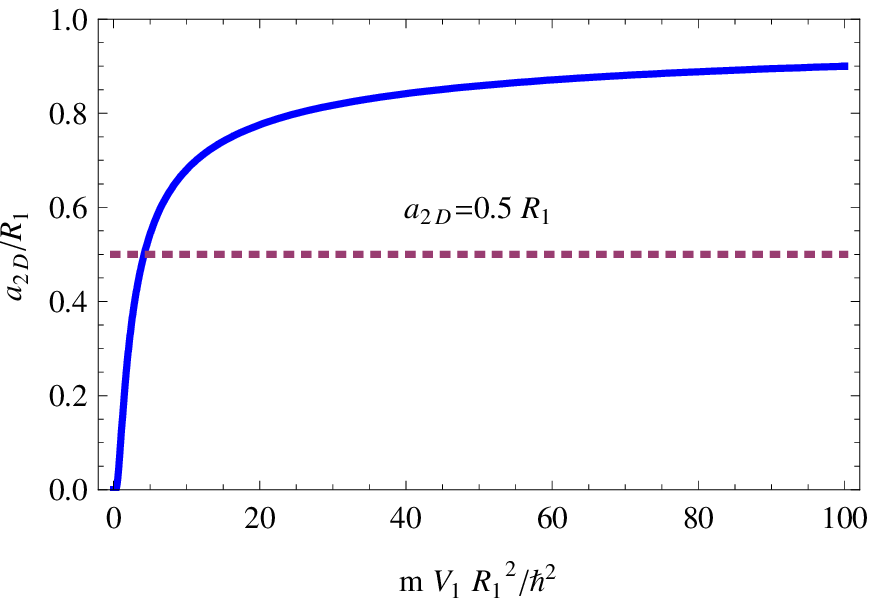} }\label{fig:a2dSD}}
\caption{(a) Scattering length of the SW potential, as a function of the depth $V_0/\varepsilon_F$, in the regime used in the simulations. The dashed line corresponds to $1/k_Fa_{2D}=1$. (b) Scattering length of the SD potential, as a function of the barrier $V_1$. The dotted line corresponds to one of the values of $R_1$ used in the simulations.}
\label{fig:a2d}       % Give a unique label
\end{figure}

For the attractive SW potential in 2D the scattering length is always non negative and diverges at null depth and at the zeros of $J_1$, corresponding to the appearance of new two-body bound states in the well. For a SW potential therefore a bound state is always present, no matter how small the attraction is. The shallow dimers have size of order $a_{2D}$ and their binding energy is given by $\varepsilon_b=-4\hbar^2/(ma_{2D}^2e^{2\gamma})$, where $\gamma\simeq0.577$ is Euler-Mascheroni's constant. A different and widely used definition of the 2D scattering length is $b=a_{2D} e^\gamma/2$, such that $\varepsilon_b=-\hbar^2/m b^2$, analogously to the 3D case. The dependence of $a_{2D}$ on the depth $V_0$ in the region where the well supports only one bound state is shown in Fig.~\ref{fig:a2dSW}; the dashed line indicates the value of scattering length at which the bare molecules have a size comparable to the mean interparticle distance $1/k_F$. The region $k_Fa_{2D}\gg 1$ corresponds to the BCS regime where interactions are weak and dimers are large and weakly bound, while $k_Fa_{2D}\ll 1$ corresponds to the BEC regime of tightly bound composite bosons. The regime in which the scattering length diverges and a bound state appears (unitarity limit)  is trivial in 2D because it corresponds to the non-interacting case; instead the resonant regime corresponds to the region $k_Fa_{2D}\sim 1$; this can be seen at the two-body level by considering the low-energy scattering amplitude $f(k)=2\pi/[\log(2/ka_{2D}e^\gamma)+i\pi/2]$~\cite{Petrov03}, which is enhanced for $k\sim 1/a_{2D}$ (with logarithmic accuracy).

For the repulsive SD potential the scattering length is always positive and smaller than $R_1$, like in 3D (see Fig.~\ref{fig:a2dSD}). Although the real physical potentials always have an attractive part, the purely repulsive models that we use can be useful in describing the cases in which the scattering length is small and positive compared to the mean interparticle distance, when it is possible to prepare metastable repulsive gas-like states without a significant production of molecules \cite{koschorreck}.

\subsection{Monte Carlo simulations}
\label{sec:qmc}
We use the fixed-node diffusion Monte Carlo (FN-DMC) method. This numerical technique solves the many-body Schroedinger equation by an imaginary time projection of an initial guess of the wavefunction. Provided that the initial guess has a finite overlap with the true ground-state, this method provides the exact energy of a systems of bosons, with a well controllable statistical error. For fermions, FN-DMC yields an upper bound for the ground-state energy of the gas, resulting from an ansatz for the nodal surface of the many-body wave function that is kept fixed during the calculation (see Refs.~\cite{QMC}). 

The fixed-node condition is enforced using an initial and guiding trial function that we choose of the standard form $\psi_T({\bf R})=\Phi_S({\bf R})\Phi_A({\bf R})$, namely the product of a purely symmetric and a purely antisymmetric term. $\Phi_A$ satisfies the fermionic antisymmetry condition and determines the nodal surface of $\psi_T$, while $\Phi_S$ is a positive function of the particle coordinates and is symmetric in the exchange of particles with equal spin (Jastrow function).  

Two opposite regimes are described by the $\Phi_A$ component. The deep attractive BEC regime, where the opposite-spin fermions are expected to pair into a condensate of dimers, can be described by an antisymmetrized product $\Phi_A({\bf R})={\cal A} \left( \phi(r_{11^\prime})\phi(r_{22^\prime})...\phi(r_{N_\uparrow N_\downarrow})\right)$ of pairwise orbitals $\phi$ corresponding to the two-body bound state of the potential $V(r)$. This wavefunction has been proposed by Leggett as a projection of the grand-canonical BCS function to a state with a finite number of particles, later extended to  the polarized case \cite{Bouchaud88} and extensively used in 3D Quantum Monte Carlo simulations \cite{QMC}. The weakly interacting regime, where a Fermi liquid description is expected to be valid, can be instead described by a typical Jastrow-Slater (JS) function with $\Phi_A({\bf R})=D_\uparrow(N_\uparrow)D_\downarrow(N_\downarrow)$, namely the product of the plane-wave Slater determinants for spin-up and spin-down particles. This description is expected to hold both in the weakly repulsive branch and in the attractive BCS regime of a weakly interacting gas where the effect of pairing on the ground-state energy is negligible. 

The symmetric part is chosen of the Jastrow form $\Phi_S({\bf R})=J_{\uparrow\downarrow}({\bf R})J_{\uparrow\uparrow}({\bf R})J_{\downarrow\downarrow}({\bf R})$. The diluteness of the gas allows us to consider just two-body Jastrow functions, with  $J_{\uparrow\downarrow}({\bf R})=\prod_{i,i^\prime}f_{\uparrow\downarrow}(r_{i,i^\prime})$, $J_{\uparrow\uparrow}({\bf R})=\prod_{i<j}f_{\uparrow\uparrow}(r_{ij})$, $J_{\downarrow\downarrow}({\bf R})=\prod_{i^\prime<j^\prime}f_{\downarrow\downarrow}(r_{i^\prime j^\prime})$, where two-body correlation functions of the interparticle distance have been introduced, which aim at reducing the statistical noise by fulfilling the cusp conditions, namely the exact behavior that is expected when two particles come close together.  In particular in the weakly interacting attractive or repulsive regimes we set $f_{\downarrow\downarrow}=f_{\uparrow\uparrow}=1$, while $f_{\uparrow\downarrow}$ is set equal to the analytical ground-state solution of the two-body Schroedinger equation in the center-of-mass frame, with the  bare two-body potentials. In the strongly attractive regime instead   $f_{\uparrow\downarrow}=1$, since the opposite-spin short-range correlation is already accounted for in the BCS orbitals; for the parallel spin correlations we use the ground-state solution of an effective two-body problem, consisting of a SD interaction with scattering length $a_{eff}=0.6 a_{2D}$, motivated by the expected interaction between molecules~\cite{Petrov03}. This choice greatly reduces the variance of the sampled energy. 

In this paper we also study the Upper Branch in the strongly attractive regime, which corresponds to a (metastable) gas of repulsive fermions; being a many-body excited state, we only perform a Variational Monte Carlo (VMC) simulation, without imaginary time projection. In this case the wavefunction is taken of the JS type, with the $f_{\uparrow\downarrow}$ functions equal to the first excited state of the two-body SW problem, aiming at enforcing orthogonality to the molecular ground-state. The nodal surface in this case arises also from the ``Jastrow'' function. This unusual non positive definite factor has already been used in the context of  3D Fermi gases for the investigation of itinerant ferromagnetism \cite{Pilati2010}.

Simulations are carried out in a square box of area $L^2=N/n$ with periodic boundary conditions (PBC). In order to fulfill those, all radially symmetric two-body functions have zero derivative at $r=L/2$, both in the Jastrow factors and in the molecular orbitals in the BCS wavefunction; this is obtained by smoothly matching each of the previously described functions to a sum of exponentials $f_e(r)=c_1+c_2[\exp(-\mu(L-r))+\exp(-\mu r)]$ at some healing length $\bar{R}$, with $\mu$ and $\bar{R}$ parameters to be optimized. The PBC also select a specific set of compatible plane-waves to be used in the JS wavefunction. It is known that finite-size effects of the JS wavefunction can be strongly depressed by considering numbers of fermions which provide a maximally symmetric Fermi surface (closed shells): we choose $N/2=13$ and  $N/2=49$.  Moreover we add a correction which can be justified with the theory of Fermi liquids (see~\cite{Ceperley1987}), namely the energy difference between the finite and infinite system is assumed to be the same as for the noninteracting case. Since this cannot assure the elimination of all finite-size effects (and in particular we do not consider the role of the effective mass), we use this correction also to assess the error-bars, on top of the statistical error. No significant finite-size effect is seen when using the BCS trial function.

\section{Energy}
\label{sec:energy}
\subsection{Weakly interacting regime}\label{subsec:weak}

The Fermi Liquid Theory (FL) prediction for the zero temperature equation of state of a weakly short-range interacting 2D gas \cite{Bloom75} is the following:
\begin{equation}\label{eq:enFL}
E_{FL}/N=E_{\text{FG}}\left[ 1+2 g+(3-4\log2)g^2\right]\;,
\end{equation}
where $E_{FG}=\hbar^2k_F^2/4m=\varepsilon_F/2$ is the energy per particle of the noninteracting gas and $g$ is the coupling constant. The peculiar logarithmic dependence of the 2D scattering amplitude on the available kinetic energy leads to an arbitrary dependence of the $2D$ coupling constant $g=1/\log(E_A/E_K)$ on a reference energy $E_K$, the $E_A$ parameter being relative to the specific choice of the potential. When finite-range effects are negligible $E_A=4\hbar^2/(ma_{2D}^2e^{2\gamma})$, which is equal to $|\varepsilon_b|$ in case of an attractive potential. Since we consider a weakly interacting Fermi liquid, we make the appropriate choice $E_K=2 \varepsilon_F$. Therefore we use the coupling constant $g=-1/\log(n a_{2D}^2 c_0)=-1/2\log(k_F b)$ with $c_0=\pi e^{2\gamma}/2$. An important remark is in order: while the coefficient of the first order term is independent of $c_0$, the reported value of the coefficient of $g^2$ depends on the chosen $c_0$, since a modification of $c_0$ gives rise to higher order contributions to the equation of state.

\begin{figure}
\centering
\resizebox{0.90\columnwidth}{!}{%
  \includegraphics{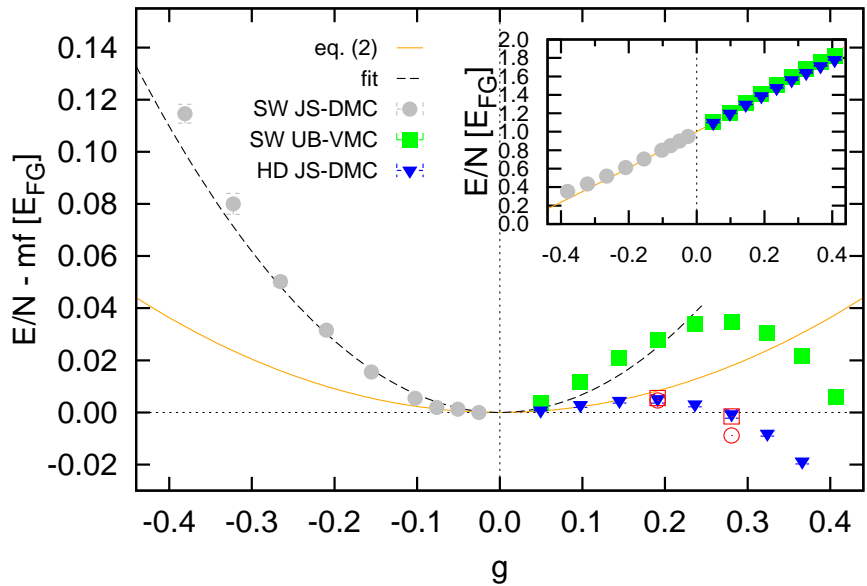} }
\caption{Energy per particle of a weakly interacting 2D Fermi gas (inset), with the mean-field contribution subtracted (main figure). The attractive regime corresponds to $g<0$, the repulsive regime to $g>0$. The results from the simulations with the SW and HD potentials are compared to the Fermi liquid theory result (\ref{eq:enFL}). The dashed line indicates the quadratic fit to the SW JS-DMC data. Empty symbols refer to calculations with the SD potential, with $R_1/a_{2D}=2,4$, only shown for two representative values of $g$.}
\label{fig:enFL}       % Give a unique label
\end{figure}

In Fig.~\ref{fig:enFL}  we show the FN-DMC results with the JS nodal surface. Negative couplings $g<0$ correspond to the ground-state in the weakly attractive regime, while positive couplings correspond to either the ground-state of the HD potential or to the Upper Branch of the attractive system. In the inset we present the energy, while in the main figure we show the energy with the mean-field result $E_{FG}(1+2g)$ subtracted, in order to emphasize beyond mean-field contributions. 
Although the three series of data obviously confirm the mean-field prediction, interesting different behaviors are found when checking the residual contributions to energy. In the $g<0$ regime we fit the coefficient $A_2$ of the second order term obtaining $A_2=0.69(4)$, to be compared to $3-4\log2=0.227$. We find $A_2\approx 3-4\log2$ only if we set $c_0=2\pi$ (as has been done in \cite{bertaina}). On the opposite side $g>0$, the Upper Branch (UB-VMC) results are approximately close to the second order fit done for the $g<0$ data, at least for small $g$. The discrepancy is due to the lack of optimization of the UB wavefunction, which is problematic since there is not a variational principle for this state. The JS results for the  SD potential with $R_1=2 a_{2D}$  are compatible with those with the HD potential. We also perform simulations with shallower SD potentials ($R_1=4 a_{2D}$), which result in departures from universality at $g\approx 0.25$.  Agreement between eq. \eqref{eq:enFL} and the HD JS-DMC results is found up to $g\approx 0.1$, which corresponds to $na^2\approx 10^{-3}$. The beyond mean-field contributions in this paramagnetic phase soon start to decrease. In a related system a ferromagnetic transition has been excluded \cite{Drummond2011}; further investigation concerning the strongly repulsive regime is however beyond the scope of this paper. The intriguing difference in the beyond mean-field terms, between the attractive case and the purely repulsive case, would also deserve further investigation, hopefully with an analytical treatment.

\subsection{BCS-BEC crossover}\label{subsec:bcs}
We now pass to discuss the main result published in \cite{bertaina}, namely the characterization of the composite bosons equation of state in the 2D BCS-BEC crossover.
The 2D mean-field BCS equations can be analytically solved \cite{Miyake83,Randeria} along the BCS-BEC crossover, in terms of the interaction coefficient $x=|\varepsilon_b|/2\varepsilon_F$. For the BCS order parameter one obtains $\Delta=2 \varepsilon_F \sqrt{x}$, while for the energy per particle one obtains $E/N=E_{FG}(1-2 x)$ and for the chemical potential $\mu=\varepsilon_F(1-x)$. For small binding energy one recovers the non interacting limit; when $x~\sim~1$, the chemical potential becomes zero and then negative, so that the role of the dimers becomes more important; for very strong binding the chemical potential of the fermions is equivalent to half the binding energy of a molecule, so the system is made of non interacting bosons. Although very useful for providing a global self-consistent picture and for setting a stringent variational upper bound to the energy per particle, the BCS solution fails in various aspects. In the BCS regime it neglects the Hartree-Fock contributions to energy (discussed in the previous Section), which are dominant, since the gap is small. In the BEC regime it misses the correct interaction energy between the bosons. In general it is not able to reproduce the logarithmic dependence of energy  on the density, which is typical of 2D.

\begin{figure}
\centering
\resizebox{0.7\columnwidth}{!}{%
  \includegraphics{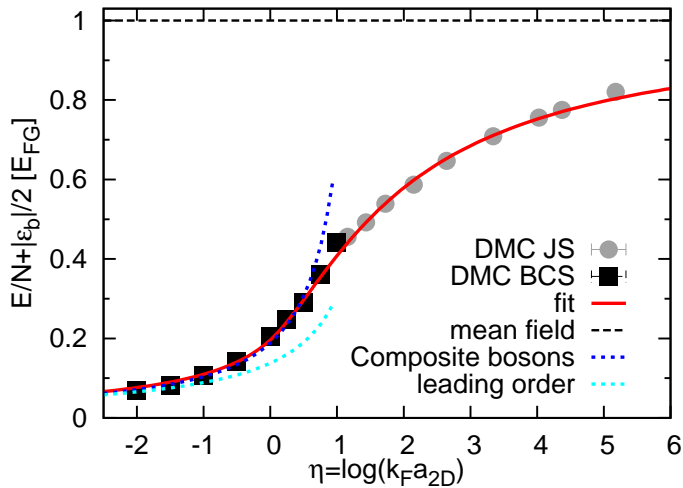} }
\caption{Energy per particle of an attractive 2D Fermi gas along the BCS-BEC crossover with the bare binding energy of the molecules subtracted. The BEC regime corresponds to $\eta\to -\infty$, the BCS regime corresponds to $\eta\to +\infty$, while the resonant regime  corresponds to $\eta\sim 1$. The various curves are described in the text. Reprint of the data in \cite{bertaina}.}
\label{fig:enBCSBEC}       % Give a unique label
\end{figure}

In Fig.~\ref{fig:enBCSBEC} we show the FN-DMC results (from \cite{bertaina}) for the equation of state of a short-range attractively interacting 2D gas as a function of the interaction parameter $\eta=\log(k_Fa_{2D})$ in units of $E_{FG}$, with $\varepsilon_b/2$ subtracted. The BCS wavefunction provides a lower energy for values of the interaction parameter $\eta\lesssim1$, while the JS function is more favorable for larger values of $\eta$.  In the deep BEC regime, besides the molecular contribution, the remaining fraction of energy corresponds to the interaction energy of the bosonic dimers. 

In the BEC regime the FN-DMC results are fitted with the equation of state of a 2D gas of composite bosons. The functional form is derived from the analysis of Beane \cite{Bose2D}, in the framework of quantum field theory; it also corresponds to previous analytical and DMC studies  (see \cite{Bose2D} and references therein). Following Beane we introduce the running coupling $g_d(\lambda)$ in terms of the scattering length of the bosons $a_d$ and the  particle density of the bosons $n_d$: $g_d(\lambda)=-1/\log{(n_d \lambda \pi^2 e^{2\gamma} a_d^2)}$,  where $\lambda$ is an arbitrary dimensionless cutoff parameter, which is present due to the truncation in the perturbative expansion; it is the analogous of the $c_0$ parameter discussed in  Subsection \ref{subsec:weak}. In the following $m_d$ is the mass of the bosons. Up to the second order in the running coupling, one can express the energy density in the following way
\begin{equation}
{\cal E}=\frac{2 \pi \hbar^2 n_d^2}{m_d} g_d(\lambda)\Bigg\lbrack 1+g_d(\lambda) \Bigl( \log{(\pi g_d(\lambda))} -\log\lambda\pi^2 +\frac{1}{2} \Bigr)\Bigg\rbrack.
\end{equation}

It is evident from the above expression that fixing the scattering length $a_d$ which appears in the definition of $g_d(\lambda)$ is not sufficient for determining the energy density if one does not declare its choice for $\lambda$, that is the form of the coupling constant. Notice again that the coefficient of the second order term does depend on the choice of $\lambda$. A convenient choice for simplifying the expression is to set $\lambda=e^{-2\gamma}/\pi^2$, so that we can introduce the coupling $g_d=-1/\log{(n_d a_d^2)}$ and we obtain
\begin{equation}
{\cal E}=\frac{2\pi\hbar^2 n_d^2}{m_d} g_d\Bigg\lbrack 1+g_d \Bigl( \log{(\pi g_d)} + 2\gamma +\frac{1}{2} \Bigr)\Bigg\rbrack.
\end{equation}

Now let us consider the case when the bosons are dimers, consisting of two paired fermions with mass $m=m_d/2$ and particle density $n=2 n_d$. There must exist a regime where the binding is so tight that the equation of state of such composite bosons is the same as in the case of point-like bosons, with the simple replacement $a_d\to \alpha a_{2D}$. Let us therefore introduce $\eta=\log{(k_F a_{2D})}$ in the previous expression,  so that the composite bosons coupling turns to be $g_d=-1/\log{(n \alpha^2 a_{2D}^2/2)}=1/(\log{4\pi}-2\eta -2\log\alpha)$. In such a situation the energy per fermion can be written in the following way:
\begin{equation}
\frac{E}{N_F}=-\frac{|\varepsilon_b|}{2}+\frac{\varepsilon_F}{2} \frac{1}{2} g_d\Bigg\lbrack 1+g_d \Bigl( \log{(\pi g_d)} + 2\gamma +\frac{1}{2} \Bigr)\Bigg\rbrack,\label{eosBose}
\end{equation}
where $-|\varepsilon_b|/2$ is the contribution from the binding energy of the dimers. From a fit of this expression we obtain $a_d=0.55(4)a_{2D}$, in agreement with the four-body calculation in Ref.~\cite{Petrov03}. In order to exemplify the importance of the second order expansion \eqref{eosBose}, in Fig. \ref{fig:enBCSBEC} we also show the leading linear contribution alone, with the same choice of the coupling constant. Without the second order term it would be impossible to determine $a_d$ with accuracy. 

To our knowledge a theoretical analysis for the running coupling of a 2D fermionic non-relativistic fluid, analogous to the one in Beane \cite{Bose2D} for the bosonic counterpart, is still lacking. It would be highly useful for interpreting the results of Subsection \ref{subsec:weak}, putting the comparison of the Quantum Monte Carlo  data and the Fermi liquid equation of state on firmer grounds.

The solid curve in Fig. \ref{fig:enBCSBEC} is a global fit of the data; the function to be fitted is a piece-wise defined function of $\eta$, the matching point being a fitting parameter and the two matched functions being the ratio of second order polynomials of $\eta$ such that the global function is continuous at the matching point up to the second derivative and the extreme regimes coincide with the known perturbative results.

\section{Contact parameter}\label{sec:tan}

The Contact parameter $C$ is a property of short-range interacting gases, measuring the amount of pairing between the particles and relating a large number of observables with each other~\cite{Tan08}. For example it can be obtained from the derivative of the equation of state with respect to the coupling parameter $C = (2\pi m/\hbar^2)d(nE/N)/d(\log k_Fa_{2D})$  or from the short-range behavior of the antiparallel pair distribution function $g_{\uparrow\downarrow}(r) \underset{r\to0}{\to} 4C/k_F^4\log^2(r/a_{2D})$~\cite{Werner} (where  $r\to0$ means $R<r\ll 1/k_F$, $R$ being the range of the potential).

\begin{figure}
\centering
\subfigure[]{\resizebox{0.40\columnwidth}{!}{%
  \includegraphics{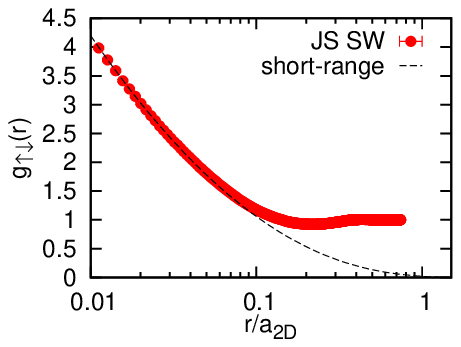} }\label{fig:grudSW}}
\subfigure[]{\resizebox{0.40\columnwidth}{!}{%
  \includegraphics{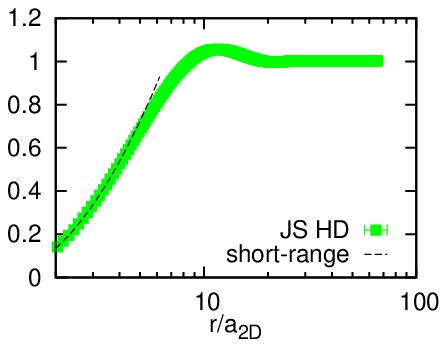} }\label{fig:grudSD}}
\caption{(a) $g_{\uparrow\downarrow}$ correlation function for the attractive gas at $\eta~=~2.15$ ($g~=~-0.25$), compared to the short-range expression (\ref{eq:grud}). (b) $g_{\uparrow\downarrow}$ correlation function for the repulsive HD gas at $g~=~0.28$.}
\label{fig:grud}       % Give a unique label
\end{figure}

We calculate the $g_{\uparrow\downarrow}$ correlation function along the crossover using FN-DMC and extract the Contact by means of the following  more refined formula, which better describes the behavior at small $r$ (see Fig. \ref{fig:grudSW}):
\begin{equation}\label{eq:grud}
 g_{\uparrow\downarrow}(r) \underset{r\to0}{\to} \frac{4C}{k_F^4}\left[-\log\left(\frac{r}{a_{2D}}\right)\left(1+\left(\frac{r}{a_{2D}e^\gamma}\right)^2\right)+\left(\frac{r}{a_{2D}e^\gamma}\right)^2\right]^2\;.
\end{equation}
The additional terms permit a more precise determination of the Contact in the deep BEC regime, where the peak in $g_{\uparrow\downarrow}$ at $r\to 0$ is very narrow since $a_{2D}$ is quite small with respect to the mean interparticle distance. 

The results for the BCS-BEC crossover have already been presented in Fig. 4 of Ref. \cite{bertaina}, where they have also been compared to the Contact extracted form the derivative of the global fit to the energy. The overall agreement between the two determinations of $C$ is a useful check of the accuracy of the trial wavefunctions used in the FN-DMC approach. Small deviations in the region $\log(k_Fa_{\text{2D}})\sim1$  point out the need of a better optimization of the trial wavefunctions, indicating that in the resonance region the JS wavefunction possesses too little pairing while the BCS wavefunction too much.

\begin{figure}
\centering
\resizebox{0.75\columnwidth}{!}{%
  \includegraphics{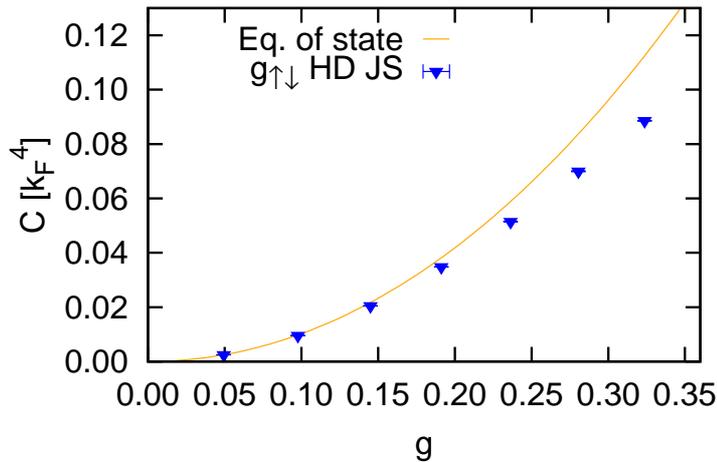} }
\caption{Contact parameter in the weakly repulsive regime. The solid line corresponds to the calculation from the derivative of the equation of state  \eqref{eq:enFL}. The triangles correspond to the extraction from the short-range behavior of $g_{\uparrow\downarrow}$ for the HD gas.}
\label{fig:contact}       % Give a unique label
\end{figure}

In Fig. \ref{fig:grudSD} we also demonstrate the determination of $C$ from the $g_{\uparrow\downarrow}$ of the repulsive HD gas. The short-range details are model dependent, nevertheless it is still possible to find an intermediate region which is universal. In this case the leading order formula for $g_{\uparrow\downarrow}$ is used, since eq. \eqref{eq:grud} is valid only for $r\ll a_{2D}$, which corresponds to the non universal region for the repulsive gas. In Fig. \ref{fig:contact} we compare $C$ extracted from $g_{\uparrow\downarrow}$ to its analytical expression obtained from eq. \eqref{eq:enFL}: $C/k_F^4=[1+(3+4\log 2)g]g^2$. Similarly to Fig. \ref{fig:enFL} deviations appear around $g=0.2$.

\section{Conclusions}
To conclude we have detailed the functional forms needed for accurately extract useful information from the FN-DMC simulations of  2D Fermi gases. Both for the energy in the weakly interacting case and in the BEC regime and for the $g_{\uparrow\downarrow}$ correlation function the knowledge of the next-to-leading order correction is crucial in order to avoid ambiguity in the measured properties. We have also shown new results on the equation of state and the Contact of  the repulsive gas in the weakly interacting regime. 

\begin{acknowledgement}
I want to acknowledge the support of the University of Trento and the INO-CNR BEC Center in Trento, where a large part of the results have been obtained, and the valuable discussions with S. Giorgini.
\end{acknowledgement}

\end{document}